\documentclass[pra,twocolumn,a4paper,showpacs,floatfix,superscriptaddress]{
revtex4}

\usepackage{latexsym}
\usepackage{amsmath}
\usepackage{amsfonts}
\usepackage{graphicx}
\usepackage{amssymb}
\usepackage{dsfont}
\usepackage{color}
\usepackage{subfigure}
\usepackage[titletoc]{appendix}

\newcommand{\ket}[1]{\ensuremath{|#1\rangle}}
\newcommand{\bra}[1]{\ensuremath{\langle #1 |}}

\newcommand{\be}{\begin{equation}}
\newcommand{\ee}{\end{equation}}
\newcommand{\mc}[1]{\ensuremath{\mathcal{#1}}}
\newcommand{\bc}{\begin{center}}
\newcommand{\ec}{\end{center}}
\newcommand{\mf}[1]{\boldsymbol{#1}}

\newcommand{\DP}{\Delta_{P}}
\newcommand{\DC}{\Delta_{C}}
\newcommand{\DPP}{\Delta_{P^{\prime}}}

\newcommand{\Gp}{R_p}

\allowdisplaybreaks[2]

\begin{document}

\title{Pulse-splitting in light propagation through $N$-type atomic media\\due
to an interplay of Kerr-nonlinearity and group velocity dispersion}

\author{Rajitha \surname{K. V.}}
\email{rajitha@iitg.ac.in} 
\affiliation{Indian Institute of Technology Guwahati, Guwahati-
781 039, Assam, India}

\author{Tarak N. \surname{Dey}}
\email{tarak.dey@iitg.ernet.in} 
\affiliation{Indian Institute of Technology Guwahati, Guwahati-
781 039, Assam, India}
\affiliation{Max-Planck-Institut f\"ur
Kernphysik, Saupfercheckweg 1, 69117 Heidelberg, Germany}

\author{J\"{o}rg \surname{Evers}}
\email{jorg.evers@mpi-hd.mpg.de} \affiliation{Max-Planck-Institut
f\"ur Kernphysik, Saupfercheckweg 1, 69117 Heidelberg, Germany}

\author{Martin \surname{Kiffner}}
\email{martin.kiffner@physics.ox.ac.uk}
\affiliation{Centre for Quantum Technologies, National University of Singapore,
3 Science Drive 2, Singapore 117543${}^1$}
\affiliation{Clarendon Laboratory, University of Oxford, Parks Road, Oxford OX1
3PU, United Kingdom${}^2$}

\begin{abstract}
We investigate the spatio-temporal evolution of a Gaussian probe pulse propagating  through a
four-level $N$-type atomic medium.
At two-photon resonance of probe-and control fields, weaker probe pulses may propagate through the medium with 
low absorption and pulse shape distortion. 
In contrast, we find that increasing the probe pulse intensity leads to a
splitting of the initially Gaussian pulse into a sequence of subpulses in the
time domain. 
The number of subpulses arising throughout the propagation can be
controlled via a suitable choice of the probe  and control field parameters. 
Employing a simple theoretical model for the nonlinear pulse propagation, we
conclude that the splitting occurs due to an interplay of Kerr nonlinearity and
group velocity dispersion. 
\end{abstract}
\maketitle

\section{\label{sec:intro}Introduction}
%
Atomic coherence and quantum interference effects are key drivers for the
control of light-matter interactions. Among the most prominent effects are
electromagnetically induced transparency (EIT)~\cite{Harris_50,Fleischhauer} and
related phenomena like the slowing and stopping of light~\cite{LVhau,kash:99,budker:99,heinze,nuclear}, the coherent storage and retrieval of 
light~\cite{hau:01,phillips:01, fleischhauer:00,dey_01}, and 
stationary light~\cite{andre:02,bajcsy:03,zimmer:06,zimmer:08,moiseev:05,moiseev:06,fleischhauer:08,lin:09,nikoghosyan:09,kiffner:10}.  
EIT can also be used to enhance the nonlinear 
susceptibility~\cite{Harris_64,wang,Fleischhauer,marangos,robert}. 
However, conventional EIT schemes are limited in their enhancement of nonlinear 
effects due to the presence of considerable absorption~\cite{Schmidt,Fleischhauer}. 
Multilevel  EIT-related atomic systems are a possible route to circumvent these limitations. 
In this spirit, Schmidt and Imam{\v o}glu  proposed 
a four-level $N$-configuration system in which the standard EIT level scheme is augmented by 
an additional off-resonantly-driven transition for the 
enhancement of the Kerr-nonlinearity~\cite{Schmidt}. This Kerr-nonlinearity 
was experimentally observed in~\cite{Kang} and has many interesting applications. 
For example, the $N$-type level system has been shown to exhibit 
cross-Kerr interactions~\cite{Sinclair} as well as the phenomena of optical bistability~\cite{Huang} and optical 
switching~\cite{sheng} in cavity systems and hollow-core optical fibres~\cite{bajcsy}. 
On the quantum level, the $N$-level system gives rise to nonlinear optical interactions that 
can be conservative~\cite{Andre} or dissipative~\cite{yamamoto}. 
These two-particle interactions were employed for the realization of strongly correlated 
polariton systems~\cite{DEChang,kiffner:10a,Hafezi}.  
From a practical point of view, it is important to measure the effect of the Kerr-nonlinearity 
on the probe field. An obvious and common way of probing an optical system is by studying its 
response to an optical pulse. In particular, the polariton systems in~\cite{DEChang,kiffner:10a,Hafezi} 
involve a sequence where an input pulse is stored, processed and then released for readout. 
An evident question is thus how the Kerr-nonlinearity in the $N$-level system modifies the 
propagation of a weak, classical probe pulse. 
However, up to now this problem has not received much attention~\cite{dey_02,Zohravi}. 

\begin{figure}[t!]
\centering\includegraphics[width=0.95\columnwidth]{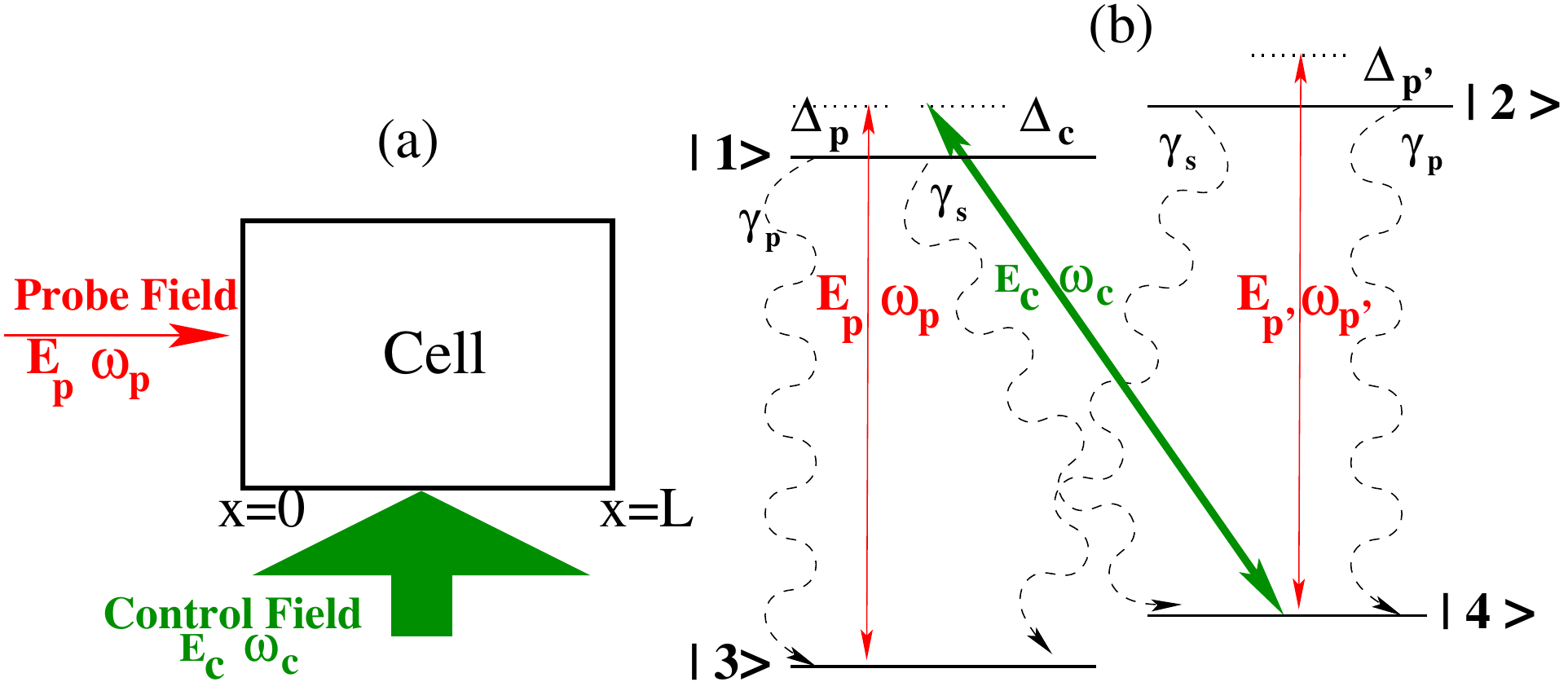}
\caption{\label{figure1}(Color online) (a) A block diagram of the system of
interest. We assume that the 
control field intensity is constant along the propagation direction. 
(b) Schematic representation of four-level atoms in $N$-type configuration.
We consider two $J=1/2$ 
Zeeman manifolds with excited states $\ket{1}=\ket{m_e=-1/2}$ and
$\ket{2}=\ket{m_e=1/2}$ and ground states 
$\ket{3}=\ket{m_g=-1/2}$ and $\ket{4}=\ket{m_g=1/2}$. 
The $\sigma^{-}$-polarized control field $\mf{E}_c$ couples to the
transition $\ket{1}\leftrightarrow\ket{4}$, 
and the $\pi$-polarized probe field $\mf{E}_p$ interacts with the
$\ket{1}\rightarrow\ket{3}$ and $\ket{2}\rightarrow\ket{4}$ transitions.}
\end{figure}

Motivated by this, here, we investigate the pulse propagation through a coherent
medium in $N$-type configuration, see Fig.~\ref{figure1}. We solve 
the Maxwell-Bloch equations governing the propagation dynamics numerically and 
consider resonant driving of the Raman transition within the $\Lambda$ sub-system. 
As a first result, we find that a  weak probe 
may propagate through the medium with low absorption and pulse shape distortion,
provided that the Kerr detuning of the probe field is sufficiently large.
However, at stronger probe pulse intensities, the numerical simulations predict
a splitting of the initially Gaussian pulse into sub-pulses, which subsequently
propagate through the medium without further splittings. To interpret these
results, we derive a simple analytical model for nonlinear pulse propagation
through the $N$-type medium. From this model, we conclude that 
the pulse splitting is a direct manifestation of 
the self-phase modulation associated with the Kerr nonlinearity in the presence of group velocity dispersion. 
The group velocity dispersion can be adjusted via the detuning of the probe and control fields with respect to 
the excited state of the $\Lambda$ sub-system. 
Measuring the temporal pulse shape of the probe pulse at the output thus indicates 
the strength of the Kerr nonlinearity experienced inside the medium. 
A similar pulse splitting phenomenon should be observable in other systems with EIT-enhanced nonlinearities giving rise to 
self-phase modulation effects.  
For example, EIT in Rydberg systems~\cite{maxwell:13} can give rise to giant photon non-linearities mediated by the 
interaction between the Rydberg atoms~\cite{gorshkov:11,hofmann:13,sevincli:11}. 
%

%
%
%
%
This  paper is organized as follows. 
In Sec.~\ref{basics} we introduce the Maxwell-Bloch equations 
governing the pulse propagation in the considered atomic four-level medium. 
In Sec.~\ref{numerics}, the propagation dynamics of the pulse is studied
numerically at different parameter conditions. In order to find a qualitative interpretation 
of our numerical findings, we introduce a model for the  description of pulse
propagation in nonlinear media in Sec.~\ref{analysis}.   
Section~\ref{conc} provides a summary and discussion of our results.
%
%
%
%
%
\section{Theoretical model \label{basics}}
We consider a  homogeneous atomic medium interacting with two laser
fields as shown in Fig.~\ref{figure1}(a). The weak probe and the strong control
fields propagate 
in perpendicular directions. 
The atomic four-level system in  $N$-type configuration is 
depicted in Fig.~\ref{figure1}(b). In the following, we assume that this level
scheme is realized 
with  two $J = 1/2$ Zeeman manifolds via polarization selection.
The weak  probe field with central frequency $\omega_p$ couples to the
$\pi$-transitions 
[$\ket{1}\leftrightarrow\ket{3}$ and $\ket{2}\leftrightarrow\ket{4}$] and is
defined as
%
  \begin{equation}
  \label{probe_field}
  \mf{E}_p (t) = \mf{e}_\pi {\cal E}_pe^{i(k_p x-\omega_{p} t)} +
\text{c.c.}, 
  \end{equation}
where $\mf{e}_\pi$ is the polarization unit vector, $k_p=\omega_p/c$ and $c$ is
the speed of light. 
The strong $\sigma^{-}$-polarized  control field with frequency $\omega_c$
couples to the
$\ket{1}\leftrightarrow\ket{4}$ transition and can be written as
%
  \begin{equation}
  \label{control_field}
  \mf{E}_c (t) = \mf{e}_{\sigma^{-}} {\cal E}_ce^{i(k_cz-\omega_{c} t)} +
\text{c.c.},~~
  \end{equation}
where $k_c=\omega_c/c$.
In electric-dipole and rotating-wave approximation, the semiclassical 
Hamiltonian of the system is 
%
 \begin{eqnarray}
 \label{Hinteraction}
 H_{\rm int}& =&
 -\hbar(\DP-\DC)\ket{4}\bra{4}-\hbar\DP\ket{1}\bra{1}\nonumber\\
 &-&\hbar(\DP+\DPP-\DC)\ket{2}\bra{2}  \nonumber  \\
 & - &\left[ (\ket{2}\bra{4}-\ket{1}\bra{3})\Omega_p
 + \ket{1}\bra{4}\Omega_c+\,\text{H.c.}\right]\,.
 \end{eqnarray}
The Rabi frequencies of the probe and control fields in 
Eq.~(\ref{Hinteraction}) are defined as 
%
 \begin{subequations}
 \label{Rabi}
 \begin{align}
 \Omega_p &= \frac{\mf{d}_{24}\cdot\mf{e}_\pi }{\hbar}{\cal E}_p\,,\\
 \Omega_c &= \frac{\mf{d}_{14}\cdot\mf{e}_{\sigma^{-}} }{\hbar}\mc{E}_c \,,
 \end{align}
 \end{subequations}
where $\mf{d}_{ij}$ is the electric dipole moment for the transition 
between levels $\ket{i}$ and $\ket{j}$. Note that $\mf{d}_{24}= -\mf{d}_{13}$ 
according to 
the Clebsch-Gordan coefficients of the considered $J = 1/2\leftrightarrow J=1/2$
level scheme. 
The detunings of probe and control fields are given by 
%
%
 \begin{subequations}
  \label{detuning}
   \begin{align}
\DP&= \omega_p -\omega_{13} \,, \\
\DPP&=\omega_p -\omega_{24}\,,\\
 \DC &= \omega_c -\omega_{14}\,. 
 \end{align}
  \end{subequations}
We model the quantum dynamics of the atoms using the master equation approach.
The matrix elements of 
the atomic density operator $\rho$ obey the following equations, 
%
 \begin{subequations}
\label{EOM}
\begin{eqnarray}
 \dot{\rho}_{11}&=&-(\gamma_p + \gamma_s)\rho_{11}-i \Omega_p
 \rho_{31}+i \Omega_p^{*}
 \rho_{13} \nonumber\\
 && + \ i \Omega_c \rho_{41}-i \Omega_c^{*} \rho_{14}~,\\
 \dot{\rho}_{12}&=&-[\gamma_p + \gamma_s-i(\DC-\DPP)]\rho_{12}+i\Omega_c
 \rho_{42}\nonumber\\
 && -\ i\Omega_p \rho_{32}-i\Omega_p^{*}\rho_{14},\\
 \dot{\rho}_{13}&=&-[(\gamma_p + \gamma_s)/2-i\DP]\rho_{13}
 +i\Omega_c\rho_{43}\nonumber\\
 &&-\ i\Omega_p(\rho_{33}-\rho_{11}),\\
 \dot{\rho}_{14}&=&-[(\gamma_p + \gamma_s)/2-i\DC]\rho_{14}
 -i\Omega_p(\rho_{34}+\rho_{12})\nonumber\\
 && +\ i\Omega_c(\rho_{44}-\rho_{11}),\\
 \dot{\rho}_{22}&=& -(\gamma_p + \gamma_s)\rho_{22}+i \Omega_p
 \rho_{42}-i \Omega_p^{*}\rho_{24}\,,\\
 \dot{\rho}_{23}&=& -\left[(\gamma_p +
\gamma_s)/2+i(\DC-(\DP+\DPP))\right]\rho_{23}\nonumber\\
 &&+\ i \Omega_p
 \rho_{43}+i \Omega_p\rho_{21}\,,\\
 \dot{\rho}_{24}&=& -\left[(\gamma_p + \gamma_s)/2-i\DPP\right]\rho_{24}+i
\Omega_p
 (\rho_{44}-\rho_{22})\nonumber\\
 &&-\ i \Omega_c\rho_{21}\,,\\
 \dot{\rho}_{33}&=& \gamma_p\rho_{11} + \gamma_s \rho_{22} -i
 \Omega_p^{*}\rho_{13}+i \Omega_p\rho_{31}\,,\\
 \dot{\rho}_{34}&=&\left[\Gamma+i(\DP-\DC)\right]\rho_{34} -i
 \Omega_p^{*}\rho_{14}\nonumber\\
 &&-\ i \Omega_p\rho_{32}-i \Omega_c\rho_{31},
 \end{eqnarray}
\end{subequations}
where $\gamma_p$ $[\gamma_s]$ is the decay rate of the $\pi$ $[\sigma]$
transition. 
These decay constants are related through the Clebsch-Gordan coefficients of the
level scheme and 
obey $\gamma_s= 2\gamma_p= 2\gamma/3$, where
$\gamma$ is the total decay rate of each excited state. 
The decay rate of the ground state coherence is denoted by $\Gamma$. 
The medium polarization induced by the probe field can be expressed in terms of 
the 
off-diagonal density matrix elements $\rho_{13}$ and $\rho_{24}$ as
%
 \begin{equation}
 \label{pol}
\mf{P} = \mc{N}( \mf{d}_{31}\rho_{13} +\mf{d}_{42}\rho_{24})\,,
 \end{equation}
where $\mc{N}$ is the number density of the medium.
The spatio-temporal evolution of the probe
pulse through the medium is governed by Maxwell's equations. 
In the  slowly varying envelope approximation,  we find
%
\begin{equation}
\label{max1}
 \left(\frac{\partial}{\partial x} + \frac{1}{c}
\frac{\partial}{\partial t}\right)\Omega_p  = i \eta
\left(\rho_{24}-\rho_{13}\right) \,,
\end{equation}
where 
\begin{align}
\eta=\gamma \frac{\mc{N} \lambda^2}{8\pi}
\label{eta}
\end{align}
is the coupling constant and $\lambda$ is the wavelength of the transition 
$\ket{1}\leftrightarrow\ket{3}$. 
%

%
\begin{figure}[t]
\centering
\subfigure[]
 {
   \includegraphics[width=0.9\columnwidth] {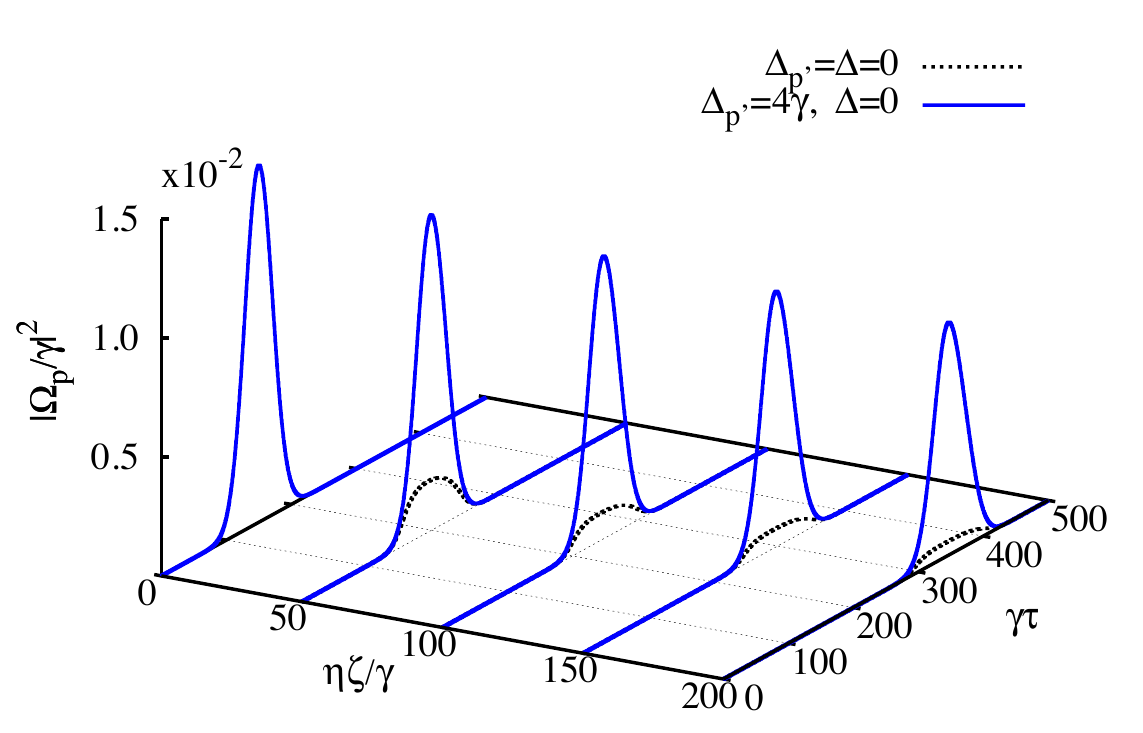}
   \label{fig:Fig2a}
 }
 \subfigure[]
 {
   \includegraphics[width=0.9\columnwidth] {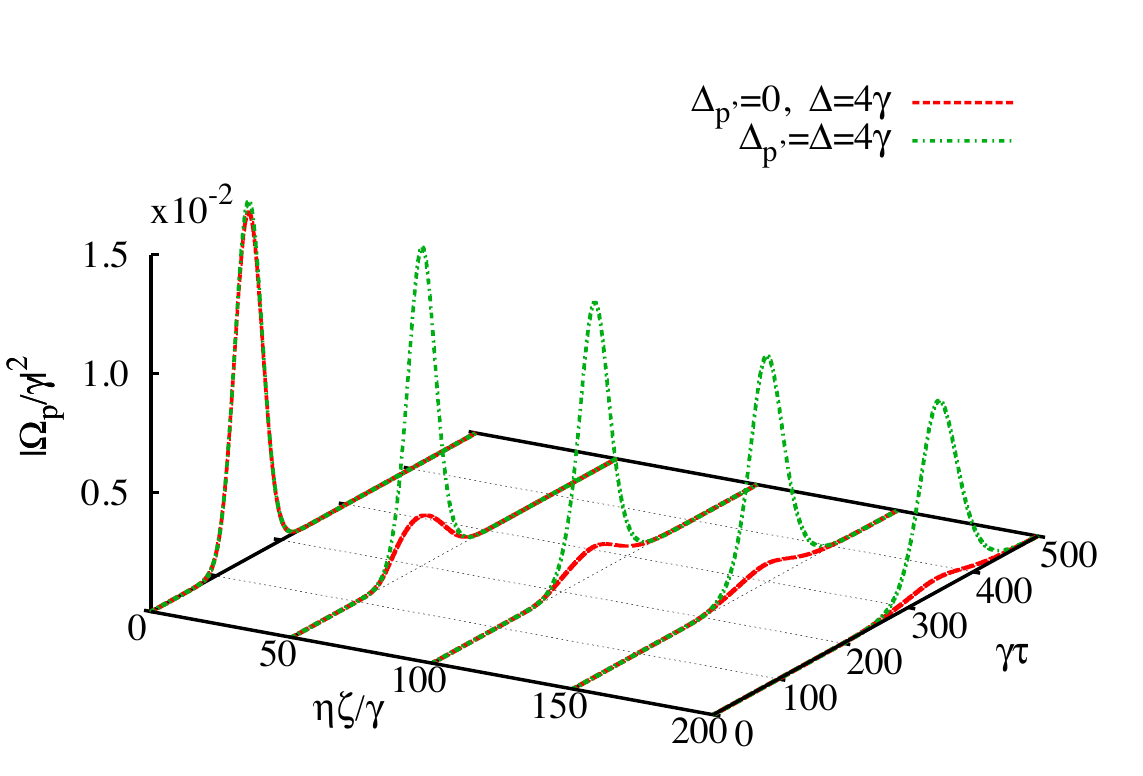}
   \label{fig:Fig2b}
 }
\caption{\label{fig:Fig2}(Color online)
Intensity $|\Omega_p/\gamma|^2$ of a weak probe pulse as a function of 
position and time for different detuning conditions. 
The  parameters in (a) and (b) are $\Omega_c=\gamma$,
$\Omega_p=\sqrt{0.015}\gamma$, $\sigma_p=30/\gamma$, and
$\Gamma=0$.
}
\end{figure}

In order to facilitate the numerical integration 
of the coupled Maxwell-Bloch equations in 
Eqs.~(\ref{EOM}) and~(\ref{max1}), we introduce a co-moving 
coordinate system 
%
\begin{equation}
\label{window}
\tau=t-\frac{x}{c},\qquad \zeta=x. 
\end{equation}

This change of coordinates reduces the partial differential equation in
Eq.~(\ref{max1}) to an 
ordinary differential equation with respect to the independent variable
$\zeta$, 
%
\begin{equation}
\label{max2}
 \frac{\partial \Omega_p}{\partial \zeta}  = i\eta
\left(\rho_{24}-\rho_{13}\right).\\
\end{equation}

Typical parameter values are
$\mc{N}=5\times10^{11}{\textrm {atoms/cm}}^3$, $\lambda= 780\ {\textrm {nm}}$,
and $\gamma= 2\pi\times10^6\ {\textrm {Hz}}$.
Throughout this paper we assume that the depletion of the strong control field
throughout its propagation can be neglected such that its intensity is
independent of position.
We further choose values of $\DP$ and $\DC$ compatible with the resonant Raman
condition $\DP=\DC$, and hence we define 
\begin{align}
\Delta=\DP=\DC \,.
\label{cond}
\end{align}

\begin{figure}[t]
\centering
   \includegraphics[width=0.85\columnwidth] {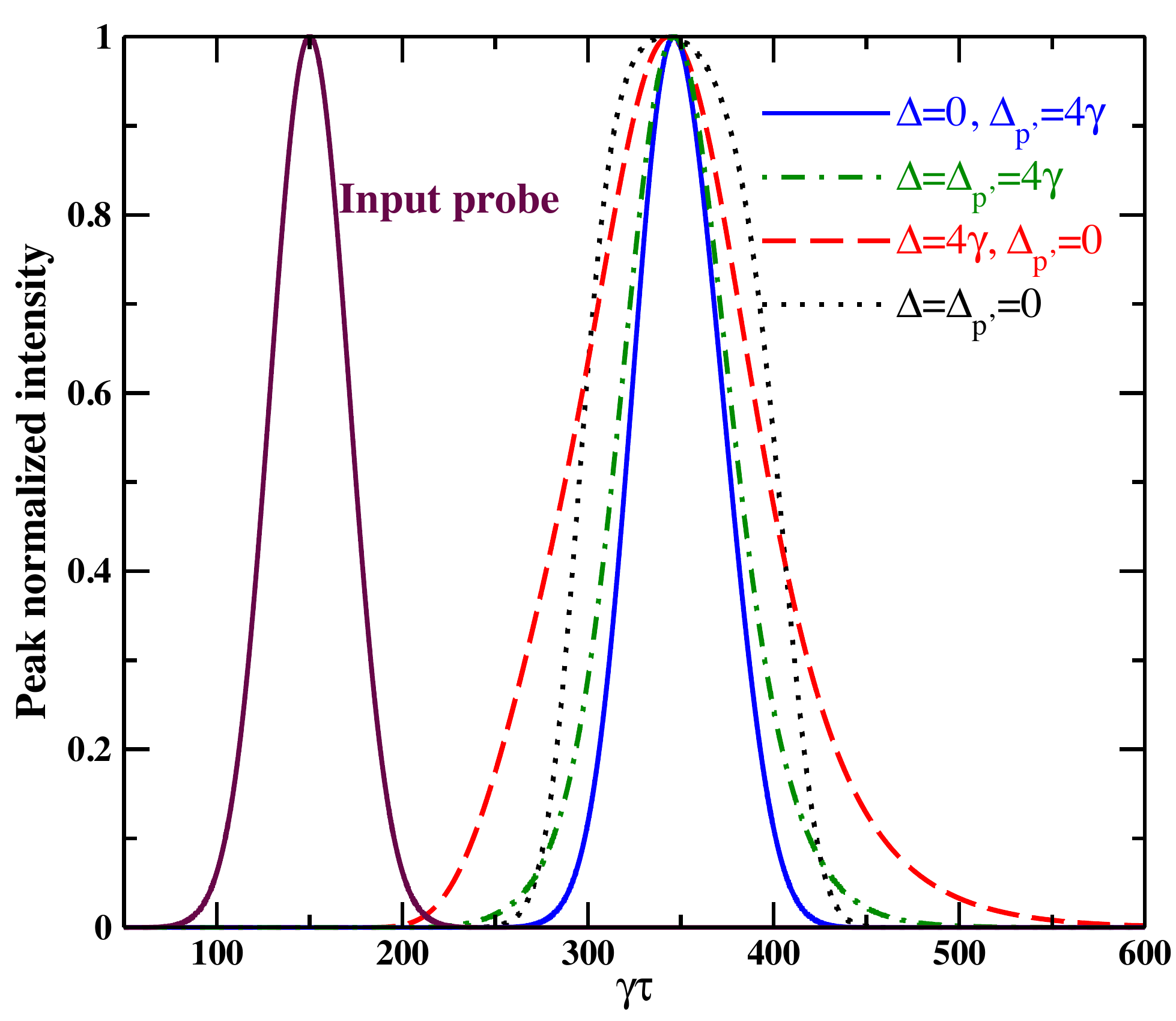}
   \caption{\label{fig:Fig3}(Color online) Probe pulse intensities at the medium
exit at $\eta\zeta/\gamma=200$. Parameters are as in Fig.~\ref{fig:Fig2}. To
facilitate the analysis of the pulse widths, all pulses are shown normalized to
their respective peak values.}
\end{figure}

%
\section{Numerical Analysis \label{numerics}}
We start our discussion with a numerical analysis of the probe field propagation
 through the atomic four-level medium. To this end, we assume that the shape of
the input pulse at the medium boundary at $\zeta=0$
is given by
%
\begin{equation}
\label{boundaryg2}
\Omega_p(\zeta=0,\tau)  = \Omega_p^0 e^{-\frac{(\tau-\tau_0)^2}{2\sigma_p^2}},  
\end{equation}
where $\Omega_p^0$ and $\sigma_p$ are the amplitude and temporal width of the
Gaussian pulse, respectively. Furthermore, all atoms are assumed to be in the
ground state $\ket 3$ initially. 

Our choice of parameters is guided by the following considerations. For suitable
parameters,
the $N$-type level scheme gives rise to self-phase modulation (SPM) due to the
optical Kerr effect~\cite{Schmidt,Agrawal_89}. As a consequence, the probe field
experiences an intensity-dependent refractive index and thus acquires a
time-dependent phase shift in accordance with its temporal intensity profile. 
This time-varying phase 
can be interpreted as a transient frequency shift. Furthermore, if the intensity
of the probe pulse is sufficiently large, the SPM gives rise to characteristic
modification of its Fourier spectrum. For example, an initially Gaussian Fourier
spectrum typically break up into several peaks in the Fourier
domain~\cite{Agrawal_89}. 

\subsection{Weak probe pulses}
In a first step we focus on weak probe pulses such that the effect of SPM is
negligible. 
To this end, we evaluate the Fourier spectra inside the medium and verify that
typical manifestations 
of SPM are absent. In this parameter regime the remaining 
system parameters governing the probe pulse propagation are given by  $\Delta$
and $\DC$. 
In the following we discuss four different detuning conditions in order to
systematically investigate 
the influence of the Raman detuning $\Delta$ and Kerr detuning $\DPP$. For this,
we evaluate all four combinations of $\Delta$ and $\DPP$ being either $0$ or
$4\gamma$:
\begin{align}
&(i) & \Delta &= 0 \quad &\textrm{and} \quad \DPP &= 4\gamma \,,\nonumber\\
&(ii) &  \Delta &= 4\gamma \quad &\textrm{and} \quad\DPP &= 4\gamma \,,
\nonumber\\
&(iii) &  \Delta &= 0\quad &\textrm{and} \quad\DPP &= 0   \,,  \nonumber\\
&(iv) & \Delta &= 4\gamma \quad &\textrm{and} \quad\DPP &= 0   \,.  \nonumber 
\end{align}

The resulting pulse propagation dynamics is shown in Fig.~\ref{fig:Fig2}. It can
be seen that $\DPP$ crucially affects the absorption. The two cases with
$\DPP=0$ lead to a transmission of less than $10\%$ of the probe pulse intensity
through the medium. In contrast, the cases $\DPP=4\gamma$ give rise to
approximately $75\%$ probe intensity transmission. This behavior can be traced
back to the absorption on the $|4\rangle \leftrightarrow |2\rangle$ transition. 
Figure~\ref{fig:Fig3} shows the effect of $\Delta$, which affects the shape and
width of the transmitted pulse, whereas its effect on the transmitted intensity
is small.

								
   %
\begin{figure}[t]
\centering
\subfigure[]
 {
   \includegraphics[width=1\columnwidth] {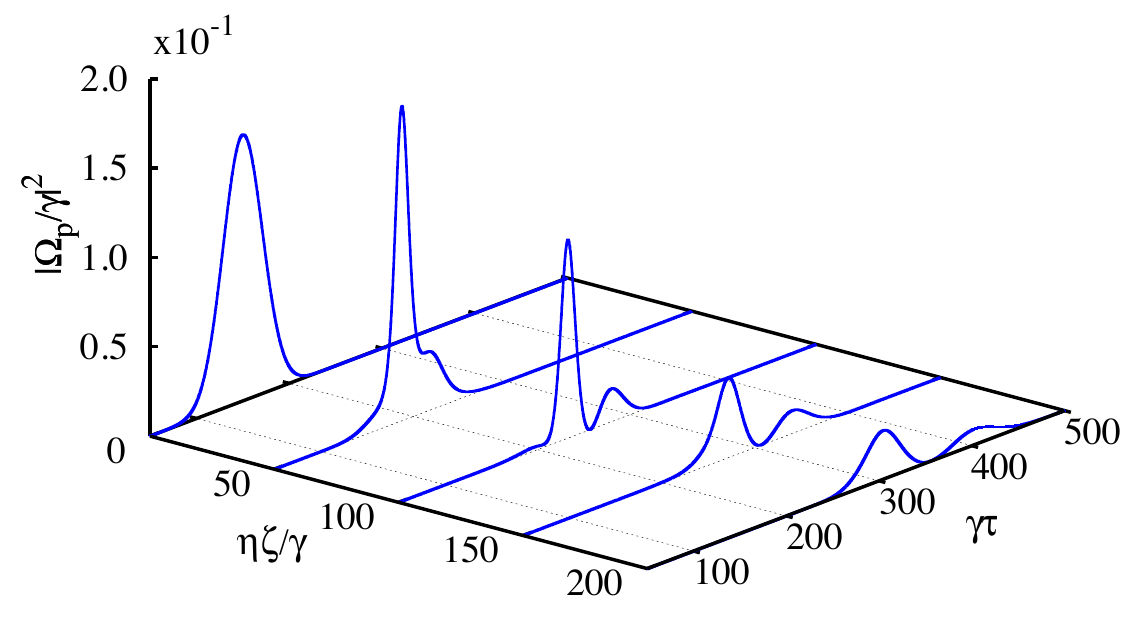}
   \label{fig:Fig4a}
 }
\subfigure[]
 {
   \includegraphics[width=1\columnwidth] {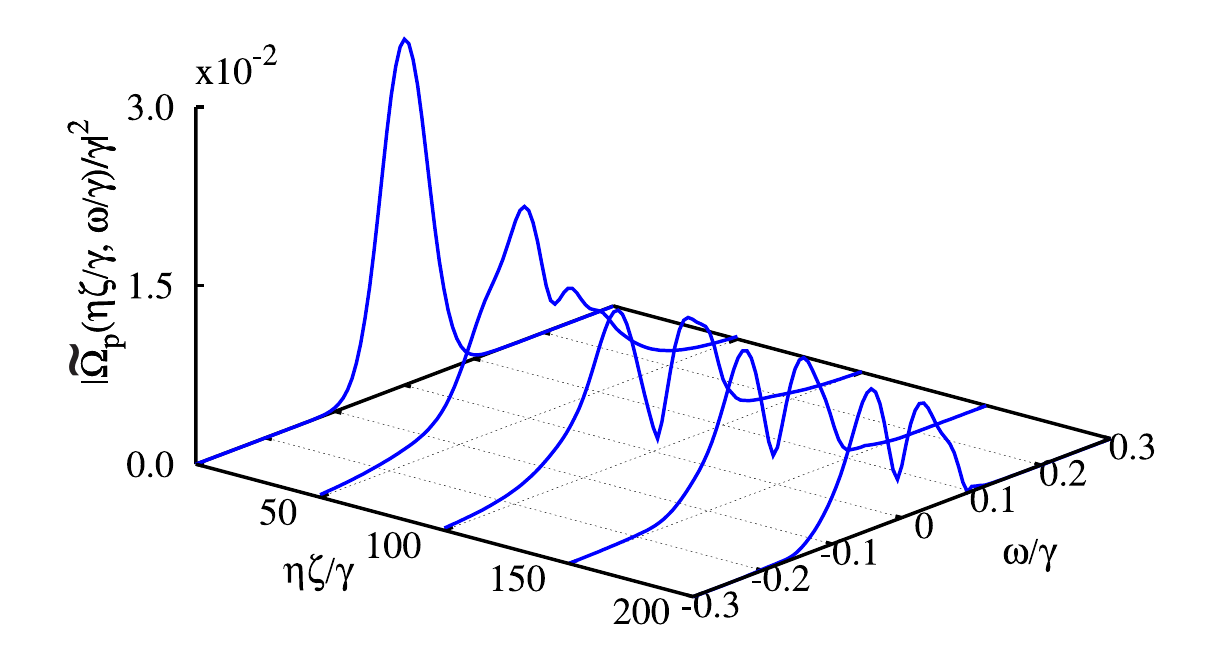}
   \label{fig:Fig4b}
 }
\caption{\label{fig:Fig4}(Color online)
(a) Intensity $|\Omega_p/\gamma|^2$ of a stronger probe pulse as function of the
propagation distance. Parameters are $\Delta= \DPP= 4\gamma$ and $\Omega_p=\sqrt
{0.15}\gamma$; all other parameters are chosen as in Fig.~\ref{fig:Fig2}. 
(b) shows the corresponding probe pulse power spectra obtained by Fourier
transformation as function of the propagation distance. 
} 
\end{figure}

\subsection{Strong probe pulses \label{strong}}
Next, we consider probe pulses with higher intensities such that the effect of
SPM 
is expected to become of relevance. Figure~\ref{fig:Fig4} shows the probe pulse
propagation dynamics as function of the propagation distance for case $(ii)$,
i.e., $\DPP = \Delta = 4\gamma$. The other parameters are chosen as in
Figs.~\ref{fig:Fig2} and \ref{fig:Fig3}, except for the intensity higher by a
factor of 10. Panel (a) shows the temporal probe pulse shape throughout the
propagation. We find that the pulse gradually breaks up, until two fully
separated pulses are obtained at $\eta\zeta/\gamma=100$.  
The shapes of these sub-pulses remain Gaussian, and they are found to
subsequently propagate through
the medium without further break-up. However, the individual subpulses broaden
as they
propagate through the medium, as it was also found in the corresponding weak
probe pulse case of Fig.~\ref{fig:Fig2}.

To further analyze this pulse break-up, Fig.~\ref{fig:Fig4b} shows the power
spectra of the probe pulse throughout its propagation, obtained by Fourier
transformation of the temporal pulse shape. The power spectrum evolves from the
Gaussian input spectrum into a double-peaked structure which is a characteristic
signature of the SPM effect. 

\begin{figure}[t!]
\centering
\includegraphics[width=1\columnwidth] {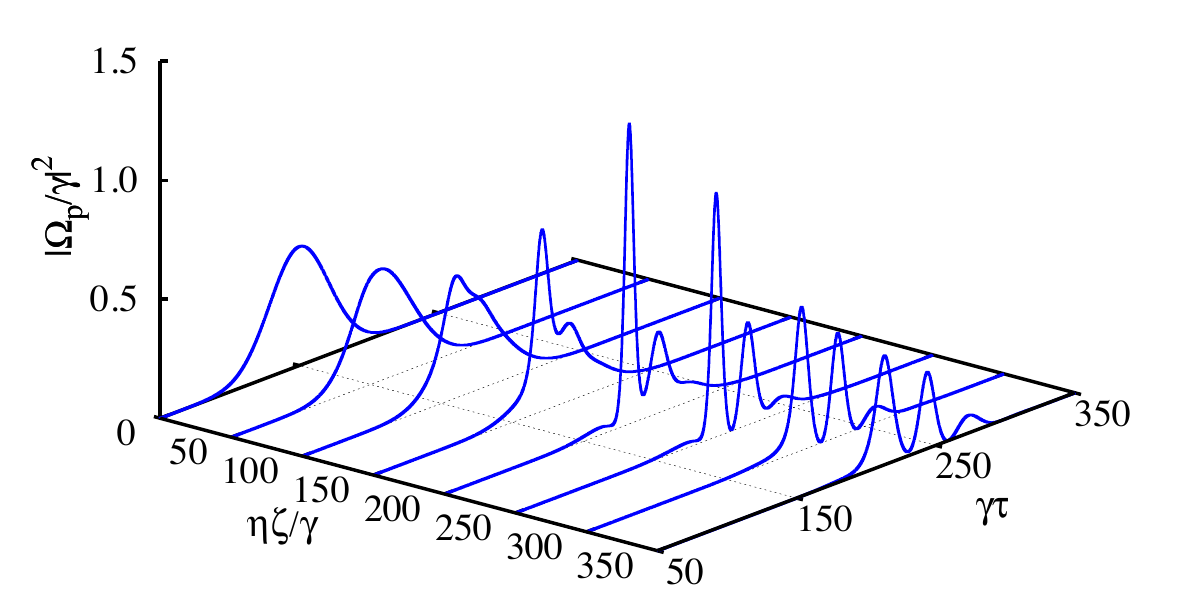}
\caption{\label{fig:Fig5}(Color online) 
Propagation dynamics of a probe pulse with 
$\Omega_p=\sqrt{0.5}\gamma$  and $\Delta=\DPP=4\gamma$. All other parameters are
chosen as in Fig.~\ref{fig:Fig2}.
}
\end{figure}

A further numerical study reveals that the number of generated subpulses depends
on the initial probe field intensity. Figure~\ref{fig:Fig5} demonstrates that a
three-peak pulse can be generated
from a single Gaussian probe pulse if the initial probe field intensity is
suitably increased. As expected, the precise  shape of the output pulse 
depends sensitively on the probe pulse intensity and the value of the 
detuning $\Delta$. For example, the three-peak shape in
Fig.~\ref{fig:Fig5} 
turns into a two-peak structure if the three-photon detuning is increased to
$\DPP=\Delta=8\gamma$. 
Finally, we verified that the shaping and splitting of the probe pulse does not
rely on the initial Gaussian pulse shape and can also be observed with
non-Gaussian initial pulse shapes, such as secant-hyperbolic pulses.

%
%
%
\section{\label{analysis} Theoretical Analysis}
Next, we provide a  theoretical interpretation of the numerical results
presented in Sec.~\ref{numerics}.  
To this end, we adapt the standard description of pulse propagation in  optical
fibers~\cite{Agrawal_89} to our system.

We find that this model reproduces  our numerical results 
accurately for weak probe pulses where the SPM phenomenon is small. For
parameters where the SPM effect becomes dominant our model allows us to gain
qualitative insights and provides an intuitive explanation for the 
pulse splitting phenomenon described in Sec.~\ref{numerics}.  

Details on our  simple model for the propagation of  optical pulses in nonlinear
media 
inspired by the standard fiber optics approach~\cite{Agrawal_89} are provided in
Appendix~\ref{Appendix}. 
There, we perform a perturbative analysis with respect to the probe field
strength to derive the  linear susceptibility $\chi(\omega)$ of the probe field,
and find that the spatio-temporal evolution of the probe pulse 
can be described by
%
 \be\label{Maxwell1}
 \left( \frac{\partial }{\partial x}+\frac{1}{c}\frac{\partial }{\partial
t}\right)\Omega_p = i\eta
\left(\sum_{n=0}^{3}\beta_n\frac{i^{n}}{n!}\frac{\partial^n
\Omega_p}{\partial t^n}\right)
 +i\eta \Gp \left|\Omega_p\right|^2\Omega_p \,.
 \ee
%
%
The parameters $\beta_n$ and $\Gp$ are determined by the  microscopic model
introduced in Sec.~\ref{basics}. In particular, 
\begin{align}
\beta_n=\left.\frac{\partial^n\chi}
  {\partial \omega^n}\right|_{\DP=\DC}^{\omega=0} 
  \label{beta}
\end{align}
is the $n$-th derivative of the medium susceptibility  $\chi(\omega)$ evaluated
at the center frequency $\omega=0$ of the probe pulse [see
Eq.~(\ref{app:beta})]. 
Note that due to the approximations used throughout the derivation, our model
described by Eq.~(\ref{Maxwell1}) is valid under the following conditions. 
($\alpha$) The probe pulse must be sufficiently long in time such that the
atomic medium follows the 
probe pulse dynamics adiabatically. 
($\beta$) The probe field spectrum must be sufficiently narrow such that 
the Taylor expansion of the linear susceptibility is justified [see
Appendix~\ref{Appendix} for details]. 
($\gamma$) The probe field intensity must be weak enough such that all
non-linear effects are well described by the 
third-order term proportional to $\Gp$.

\begin{figure}[t!]
\centering
   \includegraphics[width=1\columnwidth]{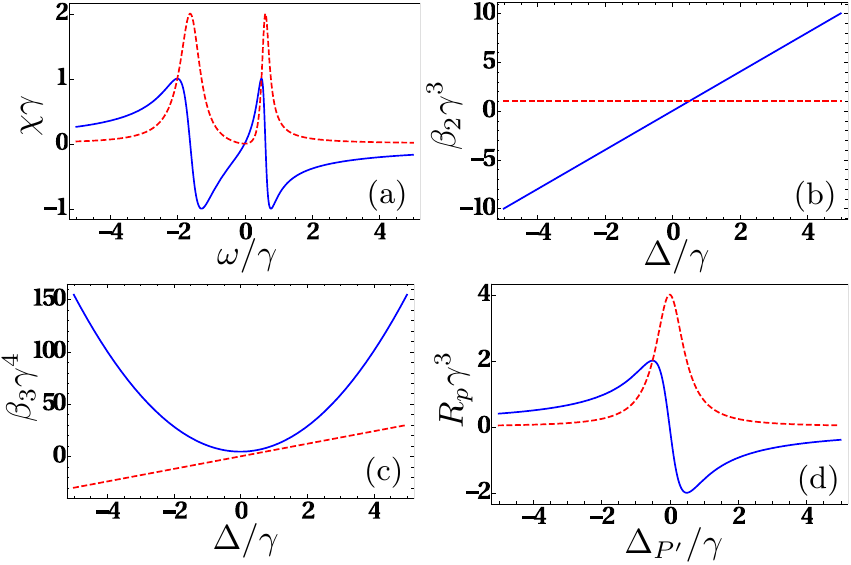}
\caption{\label{fig:Fig6}(Color online) 
(a) Real  and imaginary parts of the linear susceptibility $\chi$ in
Eq.~(\ref{Taylor_series}) 
as a function of the Fourier component $\omega$  of the probe pulse envelope. We
have 
$\beta_0=\chi(0)=0$, and $\beta_n$ is the $n$th derivative of $\chi$ at
$\omega=0$, see 
Appendix~\ref{Appendix}. The real and imaginary parts of $\beta_2$ [$\beta_3$]
are shown 
in (b) [(c)].
(d) Real and imaginary parts of  $\Gp$ as a function of 
the Kerr detuning $\Delta_{p^{'}}$ for $\Delta=4\gamma$. 
In (a)-(d), blue solid (red dashed) lines show real (imaginary) parts and 
$\Omega_c= \gamma$.
}
\end{figure}

We start by analyzing the relevant model parameters, see Fig.~\ref{fig:Fig6}.
Panel (a) shows the linear susceptibility $\chi(\omega)$ as a function of
frequency $\omega$ scanned across the probe pulse center frequency $\omega=0$ 
[see Eq.~(\ref{probe_field_freq})]. 
%
%
The parameter $\beta_0$ is the linear susceptibility $\chi$ of the probe field
evaluated at $\omega=0$, and vanishes for resonant Raman fields  [see
Eq.~(\ref{cond})]  if the ground-state decoherence rate $\Gamma$ is neglected.
$\beta_1$ is related to the group velocity of the pulse via
\begin{align}
 v_g=[1/c+\eta\text{Re}(\beta_1)]^{-1}, 
 \label{gv}
\end{align}
where Re denotes the real part. The
parameters  $\beta_2$ and  $\beta_3$ account for group velocity dispersion. The
real and imaginary parts of $\beta_2$ [$\beta_3$] are shown in
Fig.~\ref{fig:Fig6}(b) [(c)] as a function of the detuning $\Delta$. 
Both real and imaginary parts of $\beta_{2}$ and  $\beta_3$ determine the width
and amplitude
of the probe pulse  at the output. 
For $\Delta=0$ only the imaginary part of $\beta_2$ is different from zero. 
This term accounts for absorption due to 
the finite width of the EIT window in frequency space. 
For larger values of $\Delta$ the real part of $\beta_2$ increases 
and describes the change of the group velocity for frequencies away from the
central frequency of the pulse. 
In general, $\beta_3$ represents a higher-order correction to the group velocity
dispersion. 
The nonlinearity due to the Kerr transition is proportional to $\Gp$ defined in
Eq.~(\ref{zeroO}), and the 
dependence of $\Gp$ on $\DPP$  is shown in Fig.~\ref{fig:Fig6}(d). 
The imaginary part of $\Gp$ is only large for $\DPP\approx 0$, and hence
absorption on 
the Kerr transition is small for $\DPP/\gamma\gg 1$. Note that the  quantitative
impact of 
the term proportional to $\Gp$ in Eq.~(\ref{Maxwell1}) depends on the probe
pulse intensity. 
Equation~(\ref{Maxwell1}) can be numerically solved using the split-step Fourier
method~\cite{Agrawal_89}. We find excellent agreement with the numerical results
obtained in
Sec.~\ref{numerics} for parameters where the model conditions
$(\alpha)$-$(\gamma)$ are met. 
In particular, this is the case for 
all parameters where the SPM effect is small, and in this regime the 
explanation of all phenomena in terms of 
the parameters $\beta_i$ and $\Gp$ is straightforward. 
We begin with a weak probe pulse and a large detuning $\DPP$ on the Kerr
transition 
[cases (i) and (ii) in Sec.~\ref{numerics}].  
Since the imaginary part of $\Gp$ is small and hence the absorption on the Kerr 
transition is small for these parameters, 
the probe pulse can propagate through the medium without significant losses. 
The additional broadening of the probe pulse in case (ii) as compared to (i) can
be 
explained by the non-zero group velocity dispersion for the parameters in (ii). 
Next we consider  a resonant Kerr field $\DPP=0$  [cases (iii) and (iv) in
Sec.~\ref{numerics}]. 
The strong damping of the probe pulse found in Sec.~\ref{numerics}  
can be explained by the large imaginary part of $\Gp$.  The Kerr transition
hence destroys 
the EIT phenomenon and leads to absorption of the 
the probe pulse. The non-vanishing group velocity dispersion in case (iv) leads
to an additional 
broadening of the pulse as compared to (iii). 
%
\begin{figure}[t!]
\centering
\subfigure[]
 {
   \includegraphics[width=1\columnwidth] {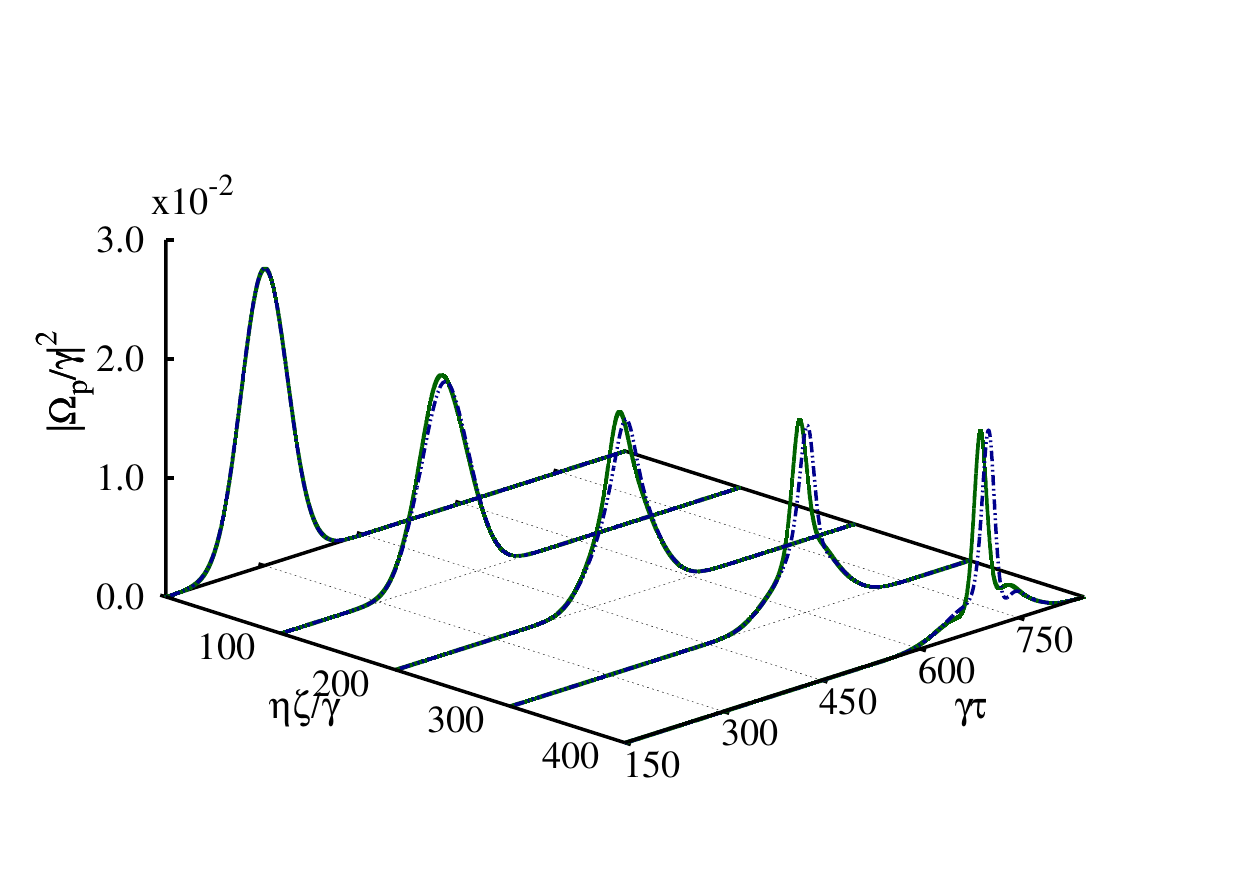}
   \label{fig:Fig7a}
   }
\subfigure[]
 {
   \includegraphics[width=1\columnwidth] {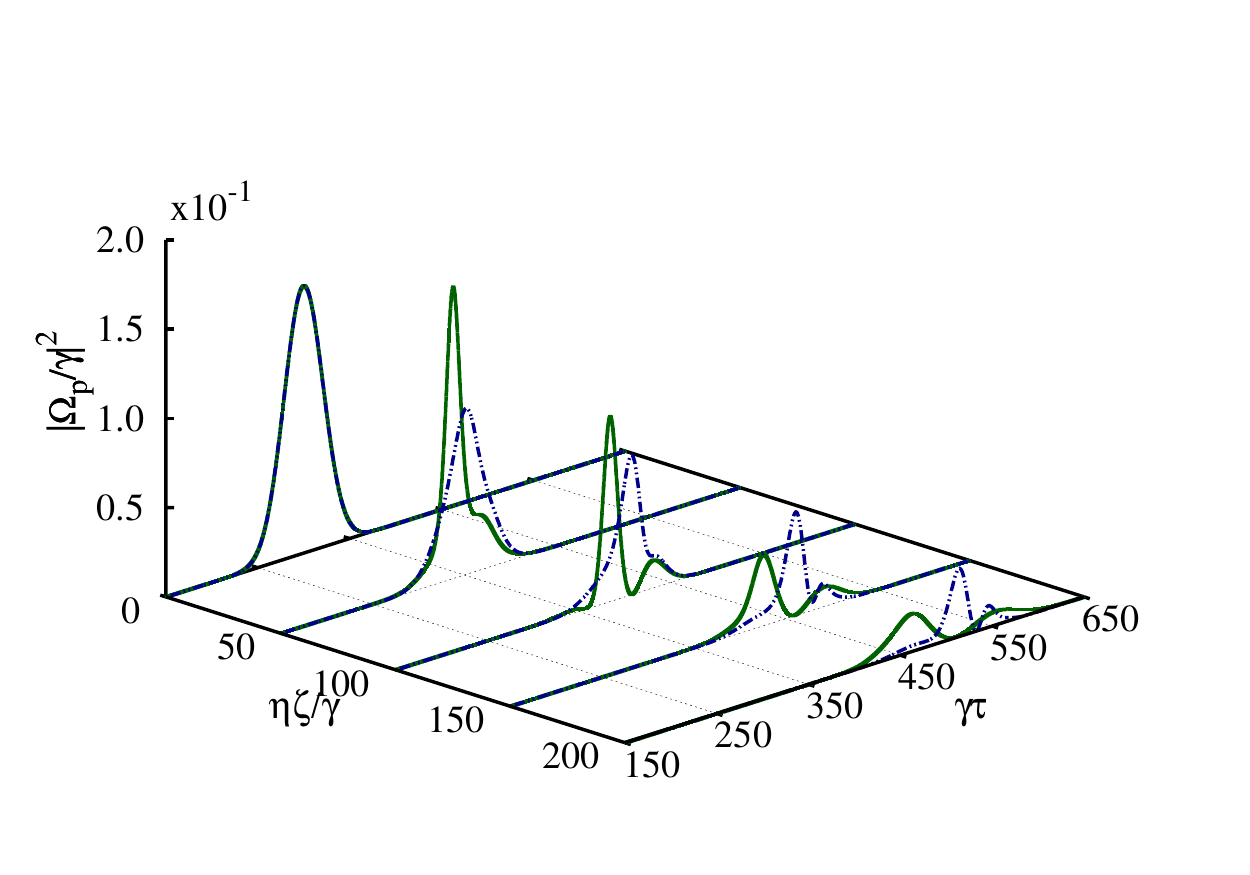}
   \label{fig:Fig7b}
 }
\caption{\label{fig:Fig7}(Color online) Comparison between the full numerical
solution to Eqs.~(\ref{EOM}) and~(\ref{max1}) 
[green solid line] and the simple model in Eq.~(\ref{Maxwell1}) [blue dot-dashed
line]. The parameters in  (a) are $\Omega_p=0.025\gamma$, $\sigma_p=50/\gamma$, 
$\Delta=\gamma$ and $\DPP=3\gamma$.. 
Results in (b) correspond to $\Omega_p=\sqrt{0.15}\gamma$,
$\sigma_p=30/\gamma$, 
and $\Delta=\DPP=4\gamma$. 
 }
\end{figure}

%
In order to gain a qualitative insight into the pulse splitting
phenomenon discussed in~Sec.\ref{strong}, we identify a parameter regime where the 
pulse splitting occurs and the conditions $(\alpha, \beta, \gamma)$ of our model are met. 
This is shown in Fig.~\ref{fig:Fig7a}, where the pulse
starts splitting at the end of the medium, and there is a reasonable agreement
between the full numerical calculations and our theoretical model. 
From the model, the pulse splitting effect can be understood as follows. 
The SPM associated with  $\Gp$ results in a splitting of the probe pulse in
frequency space. 
In the presence of group velocity dispersion described 
by $\beta_2$ and $\beta_3$, these pulses propagate at different speeds. This
explains the emergence of two different pulses in the temporal domain.

The pulse splitting as shown in Fig.~\ref{fig:Fig4}  is much more pronounced than 
in Fig.~\ref{fig:Fig7}. Our model predicts the pulse splitting effect for the parameters in Fig.~\ref{fig:Fig4},
but it is not sufficient for reproducing quantitatively correct results. 
This is shown in Fig.~\ref{fig:Fig7b}, where  we compare the theoretical model to the full
numerical simulation for parameters as in Fig.~\ref{fig:Fig4a}. The reason for the discrepancy between
the simple model in Eq.~(\ref{Maxwell1}) and the numerical approach is that the
conditions  $(\alpha)$-$(\gamma)$ for the validity of the model are 
not met for the considered parameters. In particular, the probe pulse width in
frequency space is not narrow enough to justify the adiabatic approximation
rigorously. Second, the linear susceptibility $\chi$ varies too strongly over
the width of the pulse in frequency space for $\Delta=4\gamma$ such that the
expansion of $\chi$ up to third order is insufficient.  Third, the probe pulse
intensity is not small enough such that higher order terms beyond the
third-order nonlinearity will contribute. 

Despite these shortcomings, we find that the qualitative explanation for the pulse splitting effect suggested by 
the simple model remains valid. 
To this end, we note that the power spectrum in Fig.~\ref{fig:Fig4}(b) found by numerically fourier transforming the data in 
Fig.~\ref{fig:Fig4}(a) shows a clear signature of the SPM effect. 
The initially Gaussian pulse develops into 
two separate sub-pulses with increasing propagation distance.  Let us denote the central frequencies of these pulses by 
$\omega_{+}$ and $\omega_-$, respectively. From the data in Fig.~\ref{fig:Fig4}(b) we find $\omega_{\pm}\approx\pm 0.05\gamma$. 
Next we show that the splitting of the probe pulses in 
the temporal domain is associated with the different group velocities $v_g(\omega_{\pm})$ of these two sub-pulses. 
In order to verify this, we calculate $v_g(\omega_{\pm})$ from the dispersion relation shown in Fig.~\ref{fig:Fig6}(a) 
according to Eq.~(\ref{gv}) with $\beta_1$ evaluated at $\omega=\omega_{\pm}$. As can be seen from Fig.~\ref{fig:Fig4}(b), the 
two-pulse structure starts to develop at $\eta\zeta/\gamma \approx 50$ and is fully established at $\eta\zeta/\gamma \approx 100$. We thus assume that the 
two sub-pulses evolve over a distance $\eta\Delta\zeta/\gamma \approx 130$ 
with different group velocities until the end of the medium at $\eta\zeta/\gamma = 200$ is reached. 
We estimate that the corresponding time difference is $130\gamma/\eta|1/v_g(\omega_-)-1/v_g(\omega_+)|\approx 90/\gamma$, which 
is in reasonable agreement with the time delay at $\eta\zeta/\gamma = 200$ according to Fig.~\ref{fig:Fig4}(a). 
In conclusion, we find that  the pulse splitting arises from an interplay between SPM and group velocity dispersion.

Finally, we note that the pulse splitting reduces the intensity of the individual pulses,
and suppresses nonlinear effects. Hence, the two pulses in Fig.~\ref{fig:Fig4a} do not split
again. Higher intensities such as in Fig.~\ref{fig:Fig5} lead to three pulses in the
temporal domain. Interestingly, this result does not agree with the naive expectation that the
pulse first breaks up in two pulses that subsequently split up in two pulses each, giving rise to a total
of four pulses. This outcome is a manifestation of the non-linearity of the SPM phenomenon.
%
\section{\label{conc}Conclusion} 
%
We have studied the propagation dynamics of a Gaussian pulse through an atomic
medium in four-level  $N$-type configuration under different input conditions of
control and probe field intensities and detunings. 
As our first result, we have found that at two-photon resonance condition with
large Kerr detuning, distortionless propagation of the probe pulse with low
absorption is possible. 
In contrast, when the intensity of the probe pulse becomes higher, a splitting
of the single Gaussian input pulse into a sequence of pulses may occur.
To interpret the origin of this pulse splitting, we have derived a simple
theoretical model for nonlinear pulse propagation in our model system. The
parameters of this model can be calculated from the system's steady state
density matrix. We conclude that in particular an interplay between Kerr
nonlinearity and group velocity dispersion leads to the pulse splitting. The
nonlinearity leads to a decomposition of the probe pulse  in frequency space.
Subsequently, the such formed subpulses propagate with different velocity due to
group velocity dispersion, such that a splitting in the time domain arises. 
The temporal shape of the probe pulse at the output does thus indicate 
the strength of the self-phase modulation induced by the Kerr-nonlinearity.
Our numerical analysis has shown that the number of peaks can be controlled by
adjusting the laser parameters. 
For example, the group velocity can be controlled by the detuning of the probe and control 
fields with the excited state of the $\Lambda$ sub-system. In addition, the Kerr-detuning and the laser 
intensities control the magnitude of the self-phase modulation. 

\section{Acknowledgments}
MK thanks the National Research Foundation and the Ministry of Education of
Singapore for support.
%

\appendix
\section{Model parameters\label{Appendix}} 
Here we outline the derivation of the model in Eq.~(\ref{Maxwell1}) from 
Eqs.~(\ref{EOM}) and~(\ref{max1}).  We assume 
that the probe field is sufficiently weak ($|\Omega_p|\ll \gamma,\,|\Omega_c|$) 
and pursue the perturbative approach outlined 
in~\cite{kiffner:05,kiffner:12}. The atomic density operator is expanded as 
\begin{align}
\rho = \sum\limits_{k=0}^{\infty} \rho^{(k)}, 
\label{expansion}
\end{align}
where $\rho^{(k)}$ denotes the contribution to $\rho$ in 
$k$th order in the probe field Hamiltonian 
\begin{align}
 H_{p}= - (\ket{2}\bra{4}-\ket{1}\bra{3})\Omega_p +\text{H.c.}\,.
 \label{Hp}
\end{align}
In order to obtain the  solutions $\rho^{(k)}$ we re-write the master
equation~(\ref{EOM}) as
\begin{align}
\mc{L} \rho = \mc{L}_{0}\rho -  \frac{i}{\hbar} [ H_{p} , \rho ] \,,
\label{superM}
\end{align}
where the linear super-operator $\mc{L}_{0}$ is independent of the probe field. 
Inserting the expansion~(\ref{expansion}) into Eq.~(\ref{superM}) 
leads to the following set of coupled differential equations 
\begin{align}
&\dot{\rho^{(0)}} =  \mc{L}_0 \rho^{(0)} \,, \label{rho_0} \\
&\dot{\rho^{(k)}} =  \mc{L}_0 \rho^{(k)} -\frac{i}{\hbar}[H_{p} ,\rho^{(k-1)}] ,
\quad k>0.
\label{iterative}
\end{align}
Equation~(\ref{rho_0}) describes  the interaction of 
the atom with the control field to all orders and 
in the absence of the probe field. 
Higher-order contributions to $\rho$ can be obtained 
if Eq.~(\ref{iterative}) is solved iteratively. 
Equations~(\ref{rho_0}) and~(\ref{iterative}) must be solved under the
constraints
$\text{Tr}(\rho^{(0)})=1$ and $\text{Tr}(\rho^{(k)})=0$ ($k>0$). 

First we consider the nonlinear term 
where we employ the adiabatic approximation, i.e., we assume that the temporal
length of the probe pulse  
is much longer than the lifetime of the excited states. In this case we can
find 
the nonlinear response of the medium to the probe 
field by solving for the steady state of Eqs.~(\ref{rho_0})
and~(\ref{iterative}) up to third order 
in $H_p$ and obtain 
\begin{align}
 \rho^{(0)} = \ket{3}\bra{3}\,.
\end{align}
Under the resonance condition $\Delta=\DP=\DC$ the only non-zero terms
contributing in
Eq.~(\ref{max1}) are $\rho_{24}^{(3)}$ and $\rho_{13}^{(3)}$. We find 
\begin{align}
\rho_{24}^{(3)}-\rho_{13}^{(3)}=  \Gp|\Omega_p|^2\Omega_p\,,
\label{r3}
\end{align}
where
\begin{align}
 \Gp = -\frac{2}{(\DPP+i \gamma/2)|\Omega_c|^2}\,.
 \label{zeroO}
\end{align}
Next we calculate the dominant correction  arising from the spread of the  probe
pulse in frequency 
space. To this end we consider the Fourier expansion of the slowly varying probe
pulse envelope, 
\begin{align}
  \Omega_p (t) =\int\limits_{-\infty}^{\infty} 
  \tilde{\Omega}_p(\omega) e^{-i\omega t}\text{d}\omega  \,.
  \label{probe_field_freq}
\end{align}
We solve Eqs.~(\ref{rho_0}) and~(\ref{iterative}) up to first order and obtain 
\begin{align}
 \rho^{(1)}_{13}(t) = - \int\limits_{-\infty}^{\infty}\chi(\omega)
\tilde{\Omega}_p(\omega)
 e^{-i\omega t}\text{d}\omega \,,
 \label{linorder}
\end{align}
where 
\begin{align}
 \chi(\omega) = - \frac{\omega + \DP-\DC}{(\omega + \DP+i\gamma/2)(\omega +
\DP-\DC)-|\Omega_c |^2}\,.
\end{align}
There are no other terms up to first order contributing to the coherences
$\rho_{13}$ or  $\rho_{24}$, 
and $\chi$ is shown in Fig.~\ref{fig:Fig5}. If $\chi$ is slowly varying over the
typical width of 
the probe pulse, we can expand $\chi$ in a Taylor series around $\omega=0$. For
$\DP=\DC=\Delta$ and up to third order we obtain
%
\begin{align}
\chi(\omega)\approx \sum\limits_{n=0}^{3}\frac{\omega^n}{n!}\beta_n \,,
 \label{Taylor_series}
\end{align}
where 
\begin{align}
 \label{app:beta}
\beta_n=\left.\frac{\partial^n\chi}
  {\partial \omega^n}\right|_{\DP=\DC}^{\omega=0}. 
\end{align}
The coefficients $\beta_k$ are shown in Fig.~\ref{fig:Fig6} and given by
\begin{align}
 \beta_0&= 0 \,, \\
 \beta_1&= \frac{1}{|\Omega_c|^2}\,,\\
 \beta_2 &=\frac{2(\Delta+ i\gamma/2)}{|\Omega_c|^4}\,,\\
 \beta_3 &=\frac{6[|\Omega_c|^2 +(\Delta+ i\gamma/2)^2]}{|\Omega_c|^6}\, .
\end{align}
Inserting the expansion~(\ref{Taylor_series}) into Eq.~(\ref{linorder}) gives
\begin{align}
 \rho^{(1)}_{13}(t) \approx - \sum\limits_{n=0}^3  \beta_n  \frac{i^n}{n!} 
\frac{\partial^n \Omega_p}{\partial t^n}.
\label{r1}
\end{align}
In summary, the right-hand side of Eq.~(\ref{max1}) can be written as 
\begin{align}
 i \eta (\rho_{24}-\rho_{13})\approx i\eta\left[\rho_{24}^{(3)}
-\rho_{13}^{(3)}-\rho_{13}^{(1)}\right]\,.
 \label{rhs}
\end{align}
Substituting the results from Eqs.~(\ref{r3}) and~(\ref{r1}) into
Eq.~(\ref{rhs}) allows us to obtain  Eq.~(\ref{Maxwell1}). 
%



\begin{thebibliography}{99}
\expandafter\ifx\csname natexlab\endcsname\relax\def\natexlab#1{#1}\fi
\expandafter\ifx\csname bibnamefont\endcsname\relax
  \def\bibnamefont#1{#1}\fi
\expandafter\ifx\csname bibfnamefont\endcsname\relax
  \def\bibfnamefont#1{#1}\fi
\expandafter\ifx\csname citenamefont\endcsname\relax
  \def\citenamefont#1{#1}\fi
\expandafter\ifx\csname url\endcsname\relax
  \def\url#1{\texttt{#1}}\fi
\expandafter\ifx\csname urlprefix\endcsname\relax\def\urlprefix{URL }\fi
\providecommand{\bibinfo}[2]{#2}
\providecommand{\eprint}[2][]{\url{#2}}

%
\bibitem{Fleischhauer} M. Fleischhauer, A. Imamoglu, J.P. Marangos, \rmp {\bf
77}, 633 (2005).
%
\bibitem{Harris_50} S. E. Harris, Phys. Today {\bf 50}(7), 36(1997).
%
\bibitem{LVhau} L. V. Hau, S. E. Harris, Z. Dutton and C. H. Behroozi, Nature
(London) {\bf 397}, 594 (1999).
%
\bibitem[{\citenamefont{Kash et~al.}(1999)\citenamefont{Kash, Sautenkov,
  Zibrov, Hollberg, Welch, Lukin, Rostovtsev, Fry, and Scully}}]{kash:99}
\bibinfo{author}{\bibfnamefont{M.~M.} \bibnamefont{Kash}},
  \bibinfo{author}{\bibfnamefont{V.~A.} \bibnamefont{Sautenkov}},
  \bibinfo{author}{\bibfnamefont{A.~S.} \bibnamefont{Zibrov}},
  \bibinfo{author}{\bibfnamefont{L.}~\bibnamefont{Hollberg}},
  \bibinfo{author}{\bibfnamefont{G.~R.} \bibnamefont{Welch}},
  \bibinfo{author}{\bibfnamefont{M.~D.} \bibnamefont{Lukin}},
  \bibinfo{author}{\bibfnamefont{Y.}~\bibnamefont{Rostovtsev}},
  \bibinfo{author}{\bibfnamefont{E.~S.} \bibnamefont{Fry}}, \bibnamefont{and}
  \bibinfo{author}{\bibfnamefont{M. O.}~\bibnamefont{Scully}},
  \bibinfo{journal}{Phys. Rev. Lett.} \textbf{\bibinfo{volume}{82}},
  \bibinfo{pages}{5229} (\bibinfo{year}{1999}).
%
\bibitem[{\citenamefont{Budker et~al.}(1999)\citenamefont{Budker, Kimball,
  Rochester, and Yashchuk}}]{budker:99}
\bibinfo{author}{\bibfnamefont{D.}~\bibnamefont{Budker}},
  \bibinfo{author}{\bibfnamefont{D.~F.} \bibnamefont{Kimball}},
  \bibinfo{author}{\bibfnamefont{S.~M.} \bibnamefont{Rochester}},
  \bibnamefont{and} \bibinfo{author}{\bibfnamefont{V.~V.}
  \bibnamefont{Yashchuk}}, \bibinfo{journal}{Phys. Rev. Lett.}
  \textbf{\bibinfo{volume}{83}}, \bibinfo{pages}{1767} (\bibinfo{year}{1999}).
%
\bibitem{heinze} G. Heinze, C. Hubrich, and T. Halfmann, Phys. Rev. Lett. \textbf{111}, 033601 (2013).
%
\bibitem{nuclear}K. P. Heeg, J. Haber, D. Schumacher, L. Bocklage, H.-C. Wille,
K. S. Schulze, R. Loetzsch, I. Uschmann, G. G. Paulus, R. R\"uffer, R.
R\"ohlsberger, and J. Evers, Phys. Rev. Lett. {\bf 114}, 203601 (2015).
%
\bibitem[{\citenamefont{Liu et~al.}(2001{\natexlab{a}})\citenamefont{Liu,
  Dutton, Behroozi, and Hau}}]{hau:01}
\bibinfo{author}{\bibfnamefont{C.}~\bibnamefont{Liu}},
  \bibinfo{author}{\bibfnamefont{Z.}~\bibnamefont{Dutton}},
  \bibinfo{author}{\bibfnamefont{C.~H.} \bibnamefont{Behroozi}},
  \bibnamefont{and} \bibinfo{author}{\bibfnamefont{L.~V.} \bibnamefont{Hau}},
  \bibinfo{journal}{Nature} \textbf{\bibinfo{volume}{409}},
  \bibinfo{pages}{490} (\bibinfo{year}{2001}{\natexlab{a}}).
%
\bibitem[{\citenamefont{Phillips et~al.}(2001)\citenamefont{Phillips,
  Fleischhauer, Mair, Walsworth, and Lukin}}]{phillips:01}
\bibinfo{author}{\bibfnamefont{D.~F.} \bibnamefont{Phillips}},
  \bibinfo{author}{\bibfnamefont{A.}~\bibnamefont{Fleischhauer}},
  \bibinfo{author}{\bibfnamefont{A.}~\bibnamefont{Mair}},
  \bibinfo{author}{\bibfnamefont{R.~L.} \bibnamefont{Walsworth}},
  \bibnamefont{and} \bibinfo{author}{\bibfnamefont{M.~D.} \bibnamefont{Lukin}},
  \bibinfo{journal}{Phys. Rev. Lett.} \textbf{\bibinfo{volume}{86}},
  \bibinfo{pages}{783} (\bibinfo{year}{2001}).
%
\bibitem[{\citenamefont{Dey and Agarwal}(2003)}]{dey_01}
\bibinfo{author}{\bibfnamefont{T.~N.} \bibnamefont{Dey}} \bibnamefont{and}
  \bibinfo{author}{\bibfnamefont{G.~S.} \bibnamefont{Agarwal}},
  \bibinfo{journal}{Phys. Rev. A} \textbf{\bibinfo{volume}{67}},
  \bibinfo{pages}{033813} (\bibinfo{year}{2003}).
  %
\bibitem[{\citenamefont{Fleischhauer and Lukin}(2000)}]{fleischhauer:00}
\bibinfo{author}{\bibfnamefont{M.}~\bibnamefont{Fleischhauer}}
  \bibnamefont{and} \bibinfo{author}{\bibfnamefont{M.~D.} \bibnamefont{Lukin}},
  \bibinfo{journal}{Phys. Rev. Lett.} \textbf{\bibinfo{volume}{84}},
  \bibinfo{pages}{5094} (\bibinfo{year}{2000}).
%
\bibitem[{\citenamefont{Andr\'{e} and Lukin}(2002)}]{andre:02}
\bibinfo{author}{\bibfnamefont{A.}~\bibnamefont{Andr\'{e}}} \bibnamefont{and}
  \bibinfo{author}{\bibfnamefont{M.~D.} \bibnamefont{Lukin}},
  \bibinfo{journal}{Phys. Rev. Lett.} \textbf{\bibinfo{volume}{89}},
  \bibinfo{pages}{143602} (\bibinfo{year}{2002}).
%
\bibitem[{\citenamefont{Bajcsy et~al.}(2003)\citenamefont{Bajcsy, Zibrov, and
  Lukin}}]{bajcsy:03}
\bibinfo{author}{\bibfnamefont{M.}~\bibnamefont{Bajcsy}},
  \bibinfo{author}{\bibfnamefont{A.~S.} \bibnamefont{Zibrov}},
  \bibnamefont{and} \bibinfo{author}{\bibfnamefont{M.~D.} \bibnamefont{Lukin}},
  \bibinfo{journal}{Nature} \textbf{\bibinfo{volume}{426}},
  \bibinfo{pages}{638} (\bibinfo{year}{2003}).
%
\bibitem[{\citenamefont{Zimmer et~al.}(2006)\citenamefont{Zimmer, Andr\'{e},
  Lukin, and Fleischhauer}}]{zimmer:06}
\bibinfo{author}{\bibfnamefont{F.~E.} \bibnamefont{Zimmer}},
  \bibinfo{author}{\bibfnamefont{A.}~\bibnamefont{Andr\'{e}}},
  \bibinfo{author}{\bibfnamefont{M.~D.} \bibnamefont{Lukin}}, \bibnamefont{and}
  \bibinfo{author}{\bibfnamefont{M.}~\bibnamefont{Fleischhauer}},
  \bibinfo{journal}{Opt. Commun.} \textbf{\bibinfo{volume}{264}},
  \bibinfo{pages}{441} (\bibinfo{year}{2006}).
%
\bibitem[{\citenamefont{Zimmer et~al.}(2008)\citenamefont{Zimmer, Otterbach,
  Unanyan, Shore, and Fleischhauer}}]{zimmer:08}
\bibinfo{author}{\bibfnamefont{F.~E.} \bibnamefont{Zimmer}},
  \bibinfo{author}{\bibfnamefont{J.}~\bibnamefont{Otterbach}},
  \bibinfo{author}{\bibfnamefont{R.~G.} \bibnamefont{Unanyan}},
  \bibinfo{author}{\bibfnamefont{B.~W.} \bibnamefont{Shore}}, \bibnamefont{and}
  \bibinfo{author}{\bibfnamefont{M.}~\bibnamefont{Fleischhauer}},
  \bibinfo{journal}{Phys. Rev. A} \textbf{\bibinfo{volume}{77}},
  \bibinfo{pages}{063823} (\bibinfo{year}{2008}).
%
\bibitem[{\citenamefont{Moiseev and Ham}(2005)}]{moiseev:05}
\bibinfo{author}{\bibfnamefont{S.~A.} \bibnamefont{Moiseev}} \bibnamefont{and}
  \bibinfo{author}{\bibfnamefont{B.~S.} \bibnamefont{Ham}},
  \bibinfo{journal}{Phys. Rev. A} \textbf{\bibinfo{volume}{71}},
  \bibinfo{pages}{053802} (\bibinfo{year}{2005}).
%
\bibitem[{\citenamefont{Moiseev and Ham}(2006)}]{moiseev:06}
\bibinfo{author}{\bibfnamefont{S.~A.} \bibnamefont{Moiseev}} \bibnamefont{and}
  \bibinfo{author}{\bibfnamefont{B.~S.} \bibnamefont{Ham}},
  \bibinfo{journal}{Phys. Rev. A} \textbf{\bibinfo{volume}{73}},
  \bibinfo{pages}{033812} (\bibinfo{year}{2006}).
%
\bibitem[{\citenamefont{Lin et~al.}(2009)\citenamefont{Lin, Liao, Peters, Chou,
  Wang, Cho, Kuan, and Yu}}]{lin:09}
\bibinfo{author}{\bibfnamefont{Y.-W.} \bibnamefont{Lin}},
  \bibinfo{author}{\bibfnamefont{W.-T.} \bibnamefont{Liao}},
  \bibinfo{author}{\bibfnamefont{T.}~\bibnamefont{Peters}},
  \bibinfo{author}{\bibfnamefont{H.-C.} \bibnamefont{Chou}},
  \bibinfo{author}{\bibfnamefont{J.-S.} \bibnamefont{Wang}},
  \bibinfo{author}{\bibfnamefont{H.-W.} \bibnamefont{Cho}},
  \bibinfo{author}{\bibfnamefont{P.-C.} \bibnamefont{Kuan}}, \bibnamefont{and}
  \bibinfo{author}{\bibfnamefont{I.~A.} \bibnamefont{Yu}},
  \bibinfo{journal}{Phys. Rev. Lett.} \textbf{\bibinfo{volume}{102}},
  \bibinfo{pages}{213601} (\bibinfo{year}{2009}).
%
\bibitem[{\citenamefont{Nikoghosyan and Fleischhauer}(2009)}]{nikoghosyan:09}
\bibinfo{author}{\bibfnamefont{G.}~\bibnamefont{Nikoghosyan}} \bibnamefont{and}
  \bibinfo{author}{\bibfnamefont{M.}~\bibnamefont{Fleischhauer}},
  \bibinfo{journal}{Phys. Rev. A} \textbf{\bibinfo{volume}{80}},
  \bibinfo{pages}{013818} (\bibinfo{year}{2009}).
%
\bibitem[{\citenamefont{Fleischhauer et~al.}(2008)\citenamefont{Fleischhauer,
  Otterbach, and Unanyan}}]{fleischhauer:08}
\bibinfo{author}{\bibfnamefont{M.}~\bibnamefont{Fleischhauer}},
  \bibinfo{author}{\bibfnamefont{J.}~\bibnamefont{Otterbach}},
  \bibnamefont{and} \bibinfo{author}{\bibfnamefont{R.~G.}
  \bibnamefont{Unanyan}}, \bibinfo{journal}{Phys. Rev. Lett.}
  \textbf{\bibinfo{volume}{101}}, \bibinfo{pages}{163601}
  (\bibinfo{year}{2008}).
%
\bibitem{kiffner:10} M. Kiffner and M. J. Hartmann, Phys. Rev. A {\bf 82}, 033813 (2010).
%
%
\bibitem{Harris_64} S. E.Harris, J. E. Field, A. Imamoglu, \prl {\bf 64}, 1107 (1990).
%
\bibitem{wang} H. Wang, D. Goorskey and M. Xiao, \prl {\bf 87}, 073601 (2001).
%
\bibitem{marangos} J. P. Marangos, J. Mod. Opt. {\bf 45}, 471 (1998).
%
%
%
\bibitem{robert} R. Fleischhaker and J. Evers, Phys. Rev. A {\bf 77}, 043805 (2008).
%
\bibitem{Schmidt} H. Schmidt and A. Imam{\v o}glu, \ol {\bf 21}, 1936 (1996). 
%
\bibitem{Kang} H. Kang and Y. Zhu, \prl {\bf 91}, 093601 (2003). 
%
%
\bibitem{Sinclair} G. F. Sinclair and N. Korolkova, \pra {\bf 76}, 033803 (2007).
%
%
\bibitem{Huang} J. H. Li, X. Y. Lu, J. M. Luo, and Q. J. Huang, \pra {\bf 74}, 035801 (2006).
%
%
\bibitem{sheng} J. Sheng, X. Yang, U. Khadka and M. Xiao, Opt. Exp. {\bf 19}, 17059 (2011).
%
\bibitem{bajcsy} M. Bajcsy, S. Hofferberth, V. Balic, T. Peyronel, M. Hafezi, A.
S. Zibrov, V. Vuletic, and M. D. Lukin, \prl {\bf 102}, 203902 (2009).
%
\bibitem{kiffner:10a} M. Kiffner and M. J. Hartmann, Phys. Rev. A {\bf 81}, 021806(R)
(2010).
%
\bibitem{Hafezi} M. Hafezi, D. E. Chang, V. Gritsev, E. A. Demler and M. D.
Lukin, EPL {\bf 94}, 54006 (2011).
%
\bibitem{DEChang} D. E. Chang, V. Gritsev, G. Morigi, V. Vuletic, M. D. Lukin,
and E. A. Demler, Nat. Phys. {\bf 4}, 884 (2008).
%
\bibitem{Andre} A. Andre, M. Bajcsy, A. S. Zibrov, and M. D. Lukin, \prl {\bf
94}, 063902 (2005).
%
\bibitem{Zohravi} L. E. Zohravi, M. Abedi and M. Mahmoudi, Commun. Theor. Phys.
{\bf 61}, 506 (2014).
%
%
\bibitem{dey_02} T. N. Dey and G. S. Agarwal, \pra {\bf 76}, 015802 (2007).
%
%
\bibitem{yamamoto} S. E. Harris and Y. Yamamoto, \prl {\bf 81}, 3611 (1998).
%
%
\bibitem{maxwell:13} D. Maxwell, D. J. Szwer, D. Paredes-Barato, H. Busche, J. D. Pritchard, A. Gauguet, K. J. Weatherill,
M. P. A. Jones, and C. S. Adams, Phys. Rev. Lett. \textbf{110},  103001 (2013).
%
\bibitem{sevincli:11} S. Sevincli, N. Henkel, C. Ates, and T. Pohl, \prl {\bf 107}, 153001 (2011).
%
\bibitem{hofmann:13} C. S. Hofmann, G. G\"unter, H. Schempp, M. Robert-de-Saint-Vincent, M. G\"arttner, J. Evers, 
S. Whitlock, and M. Weidem\"uller \prl {\bf 110}, 203601 (2013).
%
\bibitem{gorshkov:11} A. V. Gorshkov, J. Otterbach, M. Fleischhauer, T. Pohl, and M. D. Lukin, \prl {\bf 107}, 133602 (2011).
%
%
\bibitem{Agrawal_89} G. P. Agrawal, Nonlinear Fiber Optics (Academic, Boston, 1989).
%
\bibitem{kiffner:05} M. Kiffner and K.-P. Marzlin, \pra {\bf 71}, 033811 (2005).
%
\bibitem{kiffner:12} M. Kiffner and U. Dorner and D. Jaksch, \pra {\bf 85} 023812 (2012).
%

\end{thebibliography}
\end{document}